\documentclass[12pt]{article}

\advance \textheight by \topskip
\def\R{\mathbb R}

\def\N{\mathbb N}
\def\Z{\mathbb Z}
\def\A{\mathfrak{A}}

\def\asl{\mathfrak{sl}}

\def\su{\mathfrak{su}}
\def\u{\mathfrak{u}}

\def\cF{\mathcal F}

\def\cH{\mathcal H}
\def\I{\hbox{{1\hskip -5.8pt 1}\hskip -3.35pt I}}

\def\d{\partial}
\def\mop #1{\mathop{\sf #1}\nolimits}

\usepackage{amsmath,amssymb,euscript,graphicx}

\begin{document}
\title{
{\bf Quasi-exactly solvable models as constrained systems}}
\author{{\sf Sergey Klishevich}\thanks{E-mail:
klishevich@ihep.ru}\\
{\small {\it Institute for High Energy Physics, Protvino,
Moscow region, 142284 Russia}}}
\date{}
\maketitle

\vskip-1.0cm

\begin{abstract}
 We discuss a universal algebraic approach to quasi-exactly
 solvable models which allows us to interpret them as
 constrained Hamiltonian systems with a finite number of
 physical states. Using this approach we reproduce
 well-known two-dimensional Lie-algebraic quasi-exactly
 solvable system based on Lie algebra $\su(3)$.
\end{abstract}
\newpage

\section{Introduction}

It is well known that exactly solvable systems play very
important role in quantum theory. Unfortunately number of
such systems is quite limited. This considerably narrows
their applications. Such a situation stimulates interest to
quasi-exactly solvable systems \cite{qes, shifman, ushv1,
ushv2, kamran91, kamran93, turb&post}. In contrast to
exactly solvable models in quasi-exactly solvable systems
the spectral problem can be solved partially. Nevertheless
such systems are very interesting. Besides modeling physical
systems \cite{appl, mbody, th0105223} they can be used as an
initial point of the perturbation theory or to investigate
various nonperturbative effects~\cite{Aoyama}. Furthermore,
recently in the series of papers~\cite{nsusy} (see also
Refs.~\cite{ andrian2, klish04}) it was revealed a
connection between quasi-exactly solvable models and
supersymmetric systems with nonlinear polynomial
superalgebras~\cite{andrian}.

There exist various approaches to constructing quasi-exactly
solvable systems \cite{qes, ushv2, shifman, Zaslavskii}.
Nevertheless all of them are not universal in the sense that
they do not cover all possible quasi-exactly solvable
systems. For example, the famous Lie-algebraic approach
\cite{qes, shifman} is used to construct quasi-exactly
solvable differential operators, but does not allow, for
example, to reproduce quasi-exactly solvable systems based
on hidden dynamical symmetries with nonlinear algebras
\cite{turb&post, kamran04, klish04}. In Ref. \cite{annih}
authors presented a general construction for quasi-exactly
solvable differential operators, linear and nonlinear. But
it cannot be directly applied to quasi-exactly solvable
noncommutative systems.

Considered in this paper approach is universal because it is
formulated in terms of algebraic relations and does not
depend on any space of representation. Besides we show that
this scheme reflects a general structure of quasi-exactly
solvable systems and and their connection with constrained
systems. Therefore particularly it can be applied to
construct both usual and noncommutative quasi-exactly
solvable systems.

The paper is organized as follows. In section \ref{alg} we
formulate a universal algebraic approach to quasi-exactly
solvable systems and demonstrate its connection with
constrained systems. In section \ref{app} the algebraic
approach is used to reproduce a family of two-dimensional
quasi-exactly solvable Hamiltonians, which in the
Lie-algebraic approach are derived from a finite-dimensional
representation of the Lie algebra $\su(3)$. Brief discussion
of results is presented in section \ref{conc}.

\section{Formulation of the algebraic approach}
\label{alg}

Let us consider a set of linear operators, $A_k$ with $k=1,
\dots,n$ and $n\in\N$, on a Hilbert space $\cH$ and suppose
that they have the following commutation relations:
\begin{align}\label{algA}
 \left[A_k,\,A_l\right]&=\sum_{m=1}^nF_{klm}A_m,
\end{align}
where $F_{klm}$ are, in general, some linear
operators on $\cH$. The subspace
$\cF_A=\bigcap_{k=0}^n\mop{ker}A_k$ is the annihilator
\cite{annih} for the set of annihilating operators $A_k$.
Here we imply that the set $A_k$ is complete, i.e.
any linear operator $B$ on $\cH$ with
$\mop{ker}B\subset\cF_A$ can be represented as
$B=\sum_{k=1}^nC_kA_k$, where $C_k$ are some linear
operators on $\cH$.

Thus from \eqref{algA} it follows that the annihilating
operators $A_k$ generally form nonlinear algebra. To come to
quasi-exactly solvable systems we have to require
\begin{align}\label{fdim}
 \mop{dim}\cF_A&<\infty,
\end{align}
i.e. $\cF_A$ is a finite dimensional subspace of $\cH$.
If a Hermitian linear operator $H$ on $\cH$ has commutation
relations
\begin{align}\label{AH0}
 \left[A_k,\,H\right]&=\sum_{l=1}^nM_{kl}A_l,
\end{align}
where $M_{kl}$ are, in general, some linear operators on
$\cH$, then $H$ is a quasi-exactly solvable operator.
Indeed, the relations \eqref{AH0} imply that $H$ is an
invariant operator on finite dimensional space $\cF_A$ and
can be diagonalized on this space by a finite procedure.

It is worth noting that the annihilating operators $A_k$ can
be treated as operator constraints while $\cF_A$ can be
interpreted as a subspace of ``physical'' states. In this
context the commutation relations \eqref{AH0} can be
represented as
\begin{align}\label{AH1}
 \left[A_k,\,H\right]&\thickapprox 0
\end{align}
or even as $A_kH\thickapprox 0$.\footnote{Here a linear
operator $B\thickapprox 0$ if $By=0$ $\forall y\in \cF_A$.}
Thus, {\it a quasi-exactly solvable system with the
Hamiltonian $H$ obeying the commutation relations
\eqref{AH1} can be interpreted as a constrained Hamiltonian
system with the finite dimensional subspace of ``physical''
states $\cF_A$.}

For $n=1$ the commutation relations \eqref{algA} and
\eqref{AH0} lead to the simplest superalgebra. Indeed, the
relation $\left[A,\,H\right]\thickapprox 0$ can be
represented as
\begin{align*}
 \left[A,\,H\right]&=LA,
 \end{align*}
where $L$ is a linear operator.
This relation can be rewritten in the matrix form
\begin{align*}
 \left[Q,\,{\bf H}\right]=0,
\end{align*}
where $Q$ and $\bf H$ are matrix supercharge and
superhamiltonian:
\begin{align*}
  Q&=\begin{pmatrix}
    0&0\\A&0
  \end{pmatrix},&
  {\bf H}&=\begin{pmatrix}
    H&0\\0&L+H
  \end{pmatrix}.
\end{align*}
In one-dimensional case, $\cH={\cal C}^\omega(\R^1)$, we
have
\begin{align*}
 \left\{Q,\,Q^\dag\right\}&=P(\bf H),
\end{align*}
where the order of the polynomial $P(.)$ is equal to the
order of the annihilating operator $A$ \cite{nsusy}. Thus we
came to the supersymmetry with a nonlinear polynomial
superalgebra. Various systems with such a nonlinear
supersymmetry and their relation to quasi-exactly solvable
systems were extensively studied \cite{nsusy}.

In multi-dimensional case, $\cH={\cal C}^\omega(\R^d)$ with
$d>1$, the polynomial $P(.)$ has more involved structure and
can depend on some other operators. For example, the
two-dimensional systems with the nonlinear polynomial
supersymmetry were considered in Ref. \cite{d=2}.

The algebraic approach \eqref{algA}-\eqref{AH1} can be used
for constructing quasi-exactly solvable Hamiltonians.
Indeed, if we know the set of annihilating operators with
finite-dimensional annihilator, the corresponding
quasi-exactly solvable Hamiltonian can be constructed by
solving the commutation relations \eqref{AH0}.

Let $\cF_A$ is a finite-dimensional functional space with a
basis of linearly independent analytical functions. Such a
basis can be used to construct the corresponding set of
annihilating operators \cite{annih}. Moreover, in
Ref.~\cite{annih} the general form of a quasi-exactly
solvable operator was derived. However, in practice such a
calculation of quasi-exactly solvable Hamiltonians can be
more complicated than that based on the proposed algebraic
approach. Besides, the algebraic approach is more universal
because its formulation is not restricted by specific
realization of the Hilbert space. For example, it can be
applied to noncommutative spaces.

It is worth noting that any quasi-exactly solvable system
admits formulation in the algebraic form
\eqref{algA}-\eqref{AH1}. Indeed, by definition in any
quasi-exactly solvable system there exits a
finite-dimensional subspace, say $\cF$, which is invariant
for Hamiltonian of the system. For the finite-dimensional
subspace $\cF$ it is always possible to construct a complete
set operators $A_k$ annihilating this subspace, i.e.
$A_ky=0$ $\forall y\in\cF$ or $A_k\thickapprox 0$. Since the
Hamiltonian is invariant operator on $\cF$ we conclude that
$A_kH\thickapprox 0$ because $Hy\in\cF$ $\forall y\in\cF$.
This is equivalent to \eqref{AH1}.

Thus the general scheme of building a quasi-exactly solvable
model is the following: (1)~We choose a finite set of
independent functions, an annihilator. (2)~The
corresponding complete set of annihilating operators has to
be constructed. (3)~We calculate the
quasi-exactly solvable operator of second order using the
relations \eqref{algA}-\eqref{AH1}. Existence of such an
operator depends on the set of independent functions.
Besides, for this operator to be a Hamiltonian it has to
obey to well-known conditions~\cite{shifman, kamran93,
klish07}.

\section{Application to 2D quasi-exactly solvable systems}
\label{app}

In Ref. \cite{klish04} the algebraic approach, discussed in
the last section, was applied to the annihilator
$\cF_n=\left\{x^ky^l\,:\,0\le k\le n,\ 0\le l\le n\right\}$,
where $x,y\in\R$  and $k,l\in\Z_+$, while $n\in\N$.
The resulting Hamiltonian is equivalent to that derived from
the Lie-algebraic approach with Lie algebra
$\asl(2,\,\R)\otimes\asl(2,\,\R)$. In this section we
construct the quasi-exactly solvable Hamiltonian for the
following annihilator:
\begin{equation}\label{su3}
\cF_n=\left\{x^ky^l\,:\,0\le k+l\le n\right\}.
\end{equation}
where $x,y\in\R$  and $k,l\in\Z_+$, while $n\in\N$.
For this annihilator the corresponding annihilating
operators can be taken in the form
\begin{equation}
 A_k=\d_x^k\d_y^{n-k+1}
\end{equation}
with $k=0,\,1,\dots,n+1$.

Let us first construct general quasi-exactly solvable
operator of first order,
\begin{align}
 L = L_1(x,y)\d_x + L_2(x,y)\d_y +L_0(x,y),
\end{align}
where $L_i(x,y)$ are real-valued analytical functions.
The commutation relations
\begin{align*}
 \left[A_k,\,L\right]&\thickapprox 0,\qquad
\end{align*}
where $k=0,\,1,\dots,n+1$, lead to the following set of
differential equations:
\begin{multline}
   (l+1)(m+1)L_0^{(m,l)}(x,y)
 + (m+1)(n-k-l+1)L_2^{(m,l+1)}(x,y)\\
  {}+(l+1)(k-m)L_1^{(m+1,l)}(x,y)=0,
\end{multline}
where $-1\leq l\leq n-k+1$, $-1\leq m\leq k$, $l+m\geq 1$
and $L^{(i,j)}(x,y)=\d_x^i\d_y^jL(x,y)$.
This overdetermined system of differential equations can be
reduced to the following form:
\begin{align*}
 L_1^{(0,2)}(x,y)&=0,&\ \ \quad
 L_1^{(3,0)}(x,y)&=0,\ \ \quad&
 L_1^{(2,1)}(x,y)&=0,
\end{align*}
\vspace{-8mm}
\begin{align*}
 L_2{}^{(0,2)}(x,y) - 2L_1{}^{(1,1)}(x,y)&=0,&
 L_0{}^{(0,1)}(x,y) + nL_1{}^{(1,1)}(x,y)&=0
\end{align*}
plus the equations obtained by exchange
$x\leftrightarrow y$. This system of differential equations
has 9-parametrical solution:
\begin{align*}
 L_1(x,y)&= a_1 x + a_2 y + a_3 x^2 + a_4 xy + a_0,\\
 L_2(x,y)&= b_1x + b_2 y + a_3 xy + a_4 y^2 + b_0,\\
 L_0(x,y)&= c_0-n \left(a_3x + a_4y\right).
\end{align*}
It corresponds to the set of quasi-exactly solvable
differential operators
\begin{align}\label{J}
 J=\left\{x\d_x,\, y\d_x,\, \d_x,\, \d_y,\, x\d_y,\,
 y\d_y,\, x\left(x\d_x+y\d_y-n\right),\,
 y\left(x\d_x+y\d_y-n\right),\, \I\right\},
\end{align}
which form a representation of the Lie algebra $\u(3)$.

Now for the annihilator \eqref{su3} we construct a general
quasi-exactly solvable differential operator of second
order,
\begin{align*}
 H&=H_{11}(x,y)\d_x^2+H_{12}(x,y)\d^2_{xy}+H_{22}(x,y)\d_y^2
 + H_1(x,y)\d_x + H_2(x,y)\d_y+ H_0(x,y),
\end{align*}
where $H_{ij}(x,y)$ and $H_i(x,y)$ are real-valued
analytical functions. The commutation relations
\begin{align*}
 \left[A_k,\,H\right]&\thickapprox 0,\qquad
\end{align*}
where $k=0,\,1,\dots,n+1$, lead to the following set of
differential equations:
\begin{multline*}
     C_l^{n-k+1}C_m^kH_0^{(m,l)}
   + C_{l+1}^{n-k+1}C_m^kH_2^{(m,l+1)}\\
   + C_{l+2}^{n-k+1}C_m^kH_{22}^{(m,l+2)}
   + C_l^{n-k+1}C_{m+1}^kH_1^{(m+1,l)}\\
   + C_{l+1}^{n-k+1}C_{m+1}^kH_{12}^{(m+1,l+1)}
   + C_l^{n-k+1}C_{m+2}^kH_{11}^{(m+2,l)}=0,
\end{multline*}
where $C^k_m=\frac{k!}{(k-m)!m!}$.
This overdetermined system of differential equations can be
reduced to the following equations:
\begin{align*}
 H_{11}^{(0,3)}&=0,&
 H_{11}^{(4,1)}&=0,&
 H_{11}^{(5,0)}&=0,&
 6H_{11}^{(2,2)}-H_{22}^{(0,4)}&=0,
\end{align*}
\vskip -9mm
\begin{align*}
 3H_{11}^{(1,2)}-H_{12}^{(0,3)}=0,\qquad
 H_1^{(0,2)}+(n-1)H_{11}^{(1,2)}&=0, \\
 3H_1^{(2,0)}-6H_2^{(1,1)} - (n-1)
  \left(
    3H_{22}^{(1,2)}-H_{11}^{(3,0)}
  \right)&=0,  \\
 2H_0^{(0,1)} + n
   \left(
    2 H_1^{(1,1)}+(n-1)H_{11}^{(2,1)}
   \right)&=0, \\
  3H_{11}^{(2,1)} - 3H_{12}^{(1,2)}
   + H_{22}^{(0,3)}&=0
\end{align*}
plus the equations obtained by exchange
$x\leftrightarrow y$. This system of equations has the
following $36$-parametrical solution:
\begin{align*}
 H_{11}&=\sum_{l=0}^2\sum_{k=0}^{4-l}a_{lk}x^ky^l ,\\
 H_{22}&=\sum_{l=0}^2\sum_{k=0}^{3-l}b_{lk}x^ly^k
  + \left(a_{22}y^2 + a_{13}xy + a_{04}x^2\right)y^2,\\
 H_{12}&=\sum_{k=0}^2\left(2a_{k\,4-k}xy + a_{k\,3-k}y
  + b_{2-k\,k+1}x\right)x^{2-k}y^k
  + \sum_{k=0}^2\sum_{l=0}^{2-k}c_{kl}x^ky^l,\\
 H_1&= 2 (1-n)a_{04} x^3
   +\left((1-n)
    \left(a_{03}-b_{12}\right)+f_{11}\right)x^2
    + d_{01}x
    \\&
    + (1-n)\left(a_{21}+2 x a_{22}\right)y^2
    + \left(-2(n-1)a_{13}x^2+d_{11} x+d_{10}\right)y
    + d_{00},\\
 H_2&= 2 (1-n) a_{22}
    y^3+\left((1-n) \left(b_{03}-a_{12}\right)+d_{11}\right)
    y^2+f_{01}y
    \\&
    + (1-n)x^2\left(2a_{04}y+b_{21}\right)
    + x\left(2 (1-n)a_{13}y^2+f_{11}y+f_{10}\right)
    + f_{00},\\
 H_0&= n(n-1)a_{04} x^2
   - n\left((n-1) b_{12}+f_{11}\right)x
   + n(n-1)a_{13}xy
   \\&
   + n(n-1)a_{22}y^2
   - n\left((n-1) a_{12}+d_{11}\right)y
   + h_{00},
\end{align*}
where all the coefficients $a_{lk}$, $b_{lk}$, $c_{kl}$,
$d_{kl}$, $f_{kl}$, $h_{00}$ are real.
By direct calculation it can be shown that the resulting
Hamiltonian is equivalent to the operator
$$
 H=\sum_{\alpha,\beta=1}^9c_{\alpha\beta}J_\alpha J_\beta,
$$
where $c_{\alpha\beta}\in\R$,
which corresponds to the Lie-algebraic Hamiltonian for the
Lie algebra\footnote{In the case of the algebra
$\su(3)$ number of independent components of the matrix
$c_{\alpha\beta}$ is equal to 36.} $\su(3)$~\cite{shifman}.

For the operator $H$ to be a Hamiltonian its
coefficient functions have to obey to additional
constraints. The detail discussion of such constraints can
be found in Refs. \cite{shifman,klish04}.

\section{Conclusion}
\label{conc}

In this paper the we demonstrated that the algebraic
approach to constructing quasi-exactly solvable systems,
formulated in Ref.~\cite{klish04} can be reformulated in
terms of constrained Hamiltonian systems. This underlines
nontrivial relationship between such systems. Besides we
have shown that in the framework of this algebraic approach
one can reproduce well-known two-dimensional quasi-exactly
solvable Hamiltonian corresponding to the Lie algebra
$\su(3)$ in the Lie-algebraic approach.

In contrast to the construction of quasi-exactly solvable
differential operators, proposed in Refs.~\cite{annih,
shifman, ushv1, Zaslavskii}, considered in this paper
approach is pure algebraic and not related to specific
realization of Hilbert space, where operators live.
Therefore it is universal. For example, it can be applied to
constructing quasi-exactly solvable integral operators or
Hamiltonians on noncommutative spaces where the other
approaches do not work. We will consider construction of
such quasi-exactly solvable systems elsewhere.

Also we hope that the noted connection with constrained
Hamiltonian systems will be helpful for further development
of the theory of quasi-exactly solvable systems.


\end{document}